\documentclass[prd,a4paper,aps,showpacs,twocolumn,floats,floatfix]{revtex4}
\usepackage{epsfig}
\usepackage{graphicx}
\usepackage{color}
\usepackage{amssymb}

\newcommand{\bx}{{\mathbf x}}

\newcommand{\HH}{{\cal H}}

\newcommand{\al}{\alpha}

\newcommand{\ka}{\kappa}

\newcommand{\la}{\lambda}
\newcommand{\Om}{\Omega}

\newcommand{\be}{\begin{equation}}
\newcommand{\ee}{\end{equation}}

\newcommand{\bea}{\begin{eqnarray}}
\newcommand{\eea}{\end{eqnarray}}
\newcommand{\bean}{\begin{eqnarray*}}
\newcommand{\eean}{\end{eqnarray*}}

\newcommand{\lan}{\langle}
\newcommand{\ran}{\rangle}

\begin{document}

\title{Detection of gravitational waves from the QCD phase transition\\ with pulsar timing arrays}

\author{Chiara Caprini$^{1}$, Ruth Durrer$^{2}$ and Xavier Siemens$^3$
\\  }
\affiliation{${}^{2}$CEA, IPhT \& CNRS, URA 2306, F-91191 Gif-sur-Yvette, France \\
${}^{2}$D\'epartement de Physique Th\'eorique, Universit\'e de 
Gen\`eve\\ 24 quai Ernest 
Ansermet, CH--1211 Gen\`eve 4, Switzerland\\
${}^{3}$Center for Gravitation and Cosmology, Department of Physics, 
University of Wisconsin--Milwaukee \\ P.O. Box 413, Milwaukee, Wisconsin 53201, USA}

%\preprint{ }

\date{\today}

\begin{abstract}
If the cosmological QCD phase transition is strongly first order and lasts sufficiently long, it generates 
a background of gravitational waves which may be detected via pulsar 
timing experiments. We estimate the amplitude and the spectral shape 
of such a background and we discuss its detectability prospects. 
\end{abstract}


\maketitle

\section{Introduction}

Gravitational waves (GWs) are space-time fluctuations that propagate at the
speed of light through empty space; they were predicted by Einstein in 1916~\cite{Ei:gw}.
Because of the weakness of the gravitational interaction, GWs could provide information about
astrophysics and cosmology from regions and epochs of the Universe from which electromagnetic radiation
cannot propagate freely. For the same reason, however,  
GWs have thus far eluded direct detection, despite considerable
efforts.

Advanced configurations of existing ground-based interferometers such as LIGO~\cite{ligo} 
and VIRGO~\cite{virgo} are expected to detect GWs in the next 
years. Terrestrial interferometers have best sensitivity at a frequency $f\sim 100$~Hz, 
and are severely limited by seismic noise below a few Hertz. GWs with 
significantly lower frequency, $f \sim 10^{-9}$~Hz, are also expected to be detected by 
pulsar timing experiments in the next decade~\cite{Hobbs:2009yy,Demorest:2009ex}. A 
world-wide collaboration of astronomers, the International Pulsar Timing Array (IPTA) 
project~\cite{Hobbs:2009yy}, has been formed with the goal of detecting nano-Hz 
GWs using millisecond pulsars. Millisecond pulsars are rapidly rotating, 
highly magnetized neutron stars which emit a beam of electromagnetic radiation that sweeps over 
the earth once per rotation. They constitute extremely accurate clocks that could be
used to detect GWs. Candidates for the generation of a GW background in
the nano-Hz band are super-massive black hole binary
mergers~\cite{Wyithe:2002ep,Jaffe:2002rt,Enoki:2004ew,%
Kramer:2004hd,alberto,Sesana:2010ac} and cosmic
strings~\cite{Damour:2000wa,Damour:2001bk,Damour:2004kw,Siemens:2006yp,Olmez:2010bi}.

In this paper we study another potential candidate: the cosmological QCD phase transition, which is
believed to have taken place when the Universe had a temperature of
$T_* \simeq 100$~MeV. A GW background can be generated by the QCD phase transition 
 if it is first order, and the characteristic frequency of this background falls in the frequency band of pulsar timing experiments. 
We show here that, if the phase transition is sufficiently strong and lasts for a sufficiently long time, 
the GWs produced can be observed 
in future pulsar timing experiments. This possibility has been discussed for the first time by Witten in 
Ref.~\cite{witten}.  Here we present accurate predictions
for the spectrum of the emitted gravitational radiation as a function of the phase transition 
parameters like its temperature, strength and duration.

In the context of standard cosmology and QCD, the cosmological QCD phase 
transition is not even second order but a cross-over, and we do not expect it to 
generate GWs. However, if the neutrino chemical potential is sufficiently 
large (still well within the bounds allowed by big bang nucleosynthesis), it can become 
first order~\cite{schwarz}. Furthermore, if a sterile neutrino is the dark matter, we do 
expect a large neutrino chemical potential~\cite{misha}.

Thus, pulsar timing experiments could open a new cosmological window: 
the detection of a stochastic background of GWs could help to determine whether
the QCD phase transition is first order. The amplitude and peak
frequency of the spectrum are also sensitive to the expansion rate of the Universe 
during this phase transition~\cite{chuang}, which is currently
unconstrained.

In the next section we provide estimates for the GW 
spectrum by a first order QCD phase transition. In Section~\ref{s:pul}
we compare our results with current and expected sensitivities 
of pulsar timing arrays. We conclude in Section~\ref{s:con}.
Throughout we use the metric signature $(-,+,+,+)$ and conformal time so
that the Friedmann metric is given by 
$ds^2 =a^2(t)\left(-dt^2+ d\bx^2\right)$. The conformal Hubble
parameter is denoted by $\HH =\dot a/a = Ha$.
An over-dot denotes the derivative w.r.t conformal time $t$.

\section{Gravitational waves from a first order QCD phase transition}

Very violent processes in the early universe can lead to the
generation of GWs. One example of such violent
processes are first order phase transitions \cite{witten,hogan83,hogan86,TW,beta}, which can lead to
GW production via the collision of bubbles of the true
vacuum \cite{Kos92,Kos93,KKT,apreda,apreda2,HuKo2,HuKo,bubble,CRTG} and via the turbulence and magnetic fields they
can induce in the cosmic plasma \cite{hogan832,Kos02,dolgov,CR,gogob,THelical,THelical2,CRG}.

The GWs generated by a source are determined by
 the linearized Einstein equation for tensor perturbations in a
Friedmann background~\cite{mybook},
\be
\ddot h_\pm + 2\frac{\dot a}{a}\dot h_\pm +k^2h_\pm = 8\pi G a^2\rho\Pi_\pm \; .
\ee
Here $\rho$ is the background cosmological energy density, $a$ is the
scale factor, $\Pi_\pm$ are the two tensor helicity modes of the
(dimensionless) anisotropic stress which is the source of GWs, and $h_\pm$ are the helicity modes of the
GW.  If an anisotropic stress is generated at some time $t_*$ in the radiation dominated era, from a source
with relative energy density $\Om_{S*}=\rho_{S*}/\rho_{*}$, we expect
$\Pi$ to be at best of the order of $\Om_{S*}$. The GW energy density (from two polarizations $\pm$
which contribute equally) is given by 
\be
\rho_{\rm GW}(t) = \frac{\lan\dot h_+(\bx,t)\dot h_+(\bx,t)
  \ran}{8 \pi Ga^2(t)} \;.  
\ee 
Due to statistical homogeneity, $\rho_{\rm GW}$ is independent of the position. The
GW energy spectrum per logarithmic unit of frequency
$\frac{d\rho_{\rm
    GW}(k,t)}{d\log(k)}$ is defined by
$$
\rho_{\rm GW} = \int\frac{dk}{k}\frac{d\rho_{\rm GW}(k,t)}{d\log(k)} \;.
$$
Detailed semi-analytical and numerical calculations have been performed in the past in order to calculate the
GW energy spectrum from first order phase transitions \cite{Kos92,Kos93,KKT,apreda,apreda2,HuKo2,HuKo,bubble,CRTG,hogan832,Kos02,dolgov,%
CR,gogob,THelical,THelical2,CRG}.
In this paper we simply use analytic fits to the most recent results, as presented in the following.

Concerning the GW signal from bubble collisions, we use the shape of the spectrum 
proposed in Ref.~\cite{CRTG}, but rescale the amplitude to agree
with the numerical result of Ref.~\cite{HuKo}. As a result, the GW
energy density emitted by this source is well approximated by
\begin{center}
Bubble collisions:
\end{center}
\be
\label{e:Ombub}
\frac{d\Om^{({\rm B})}_{\rm GW}h^2}{d\log k} \simeq \frac{2}{3\pi^2} \,
h^2\Om_{r0} \left(\frac{\HH_*}{\beta}\right)^2\Om_{S*}^2v^3
\frac{\left(k/\beta\right)^3}{1+
   \left(k/\beta\right)^4} \;. 
\ee
Here $\Om_{r0}$ denotes the radiation energy density today, 
$h=H_0/(100$km/s/Mpc) is the present Hubble parameter in units of 100km/s/Mpc, $\beta^{-1}$ is the duration of the phase
transition, $v$ is the expansion velocity of
the bubbles, and $k$ is the co-moving wave-number or frequency of the GW. 
The GW spectrum is proportional to the relative energy density in the source $\Om_{S*}^2$, to the ratio between the duration of the phase transition and the Hubble time $(\HH_*/\beta)^2$, and to the bubble velocity $v^3$ \cite{HuKo}. 
 
The QCD phase transition is expected to happen at the temperature
$T_*\simeq 100$ MeV, when the kinetic and magnetic Reynolds numbers of the cosmic fluid are
very large~\cite{CRG}. The bubbles which rapidly expand and collide are therefore expected to generate magneto-hydrodynamical
(MHD) turbulence in the cosmic fluid. The kinetic energy of the turbulent
motions and the magnetic fields sustained by the MHD turbulence also
induce GWs: Ref.~\cite{CRG} presents the latest 
semi-analytical calculation of the GW spectrum from MHD turbulence. There are two important
differences with respect to the GW signal from bubbles. First,
turbulence lasts beyond the duration of the phase transition: this
leads to an enhancement of the signal on large (super-horizon) scales
\cite{CRG}. Second, the time correlation properties of the
anisotropic stress source are different. For bubble collisions, the
source is totally coherent (see~\cite{CRTG,CRG}), while for MHD
turbulence the source is coherent only over about one characteristic
wavelength \cite{CRG}. This leads to a difference in the peak position of the
GWs from the two sources: while the signal from
bubble collisions peaks at $k_p\sim \beta$, the inverse duration of
the phase transition, the peak of the MHD signal is related to the
bubble size: the peak wavelength becomes therefore $\la_p\sim
R_*\simeq v/\beta$.  The analysis of Ref.~\cite{CRG} finds a peak at
about $k_p\sim \pi^2\beta/v$. We can fit the GW
spectrum obtained in \cite{CRG} by the following formula:
\begin{center}
MHD turbulence:
\end{center}
\bea
\frac{d\Om^{({\rm MHD})}_{\rm GW}h^2}{d\log k} &\simeq& \frac{8}{\pi^6} \, h^2\Om_{r0}\,
\frac{\HH_*}{\beta} \, \Om_{S*}^{3/2} \, v^4  \nonumber\\
&& \hspace{-2cm} \times\frac{\left(k/\beta \right)^3}{\left(1+4k/\HH_*\right)\left[
  1+ (v/\pi^2)(k/\beta)\right]^{11/3}} \;.   \label{e:Omturb}
\eea

For large scales, $k\ll k_p$, both spectra in Eqs. (\ref{e:Ombub}) and (\ref{e:Omturb}) increase as $k^3$: this
behavior is simply due to causality~\cite{CR,RC}. Since the
anisotropic stresses are generated by a causal process, their spectrum
is white noise at scales larger than the typical correlation scale of
the source, which corresponds to the bubble size.  The white noise
spectrum is inherited by the GWs: $\lan |\dot h|^2\ran
\propto $ const.~, so that the GW energy density scales
simply with the phase space volume $k^3$. The behavior on small
scales, $k\gg k_p$, depends on the source power spectrum and on the unequal
time correlation properties of the source, see Ref.~\cite{R}.  In particular, the result of Eq. (\ref{e:Ombub}) resides on the assumption 
that the bubbles are infinitely thin: this assumption holds if the bubbles propagate as detonations and causes the $k^{-1}$ slope at high wave-numbers \cite{HuKo,CRTG}. On the other hand, the $k^{-5/3}$ decay of Eq.~(\ref{e:Omturb}) is a consequence of the Kolmogorov type spectrum assumed for the MHD turbulent motions at high wave-numbers. In addition, the
slope of the MHD signal changes at sub-horizon scales, $\HH_*<k<k_p$,
from $k^3$ to $k^2$ due to the long duration of the source (c.f.
\cite{CRG}).

From the the above formulas for the GW spectra we see that the basic
ingredients which determine the peak position and amplitude are simply
the fractional energy density of the source $\Om_{S*}$, the duration
of the phase transition $\beta^{-1}$, and the bubble velocity $v$
(besides obviously the temperature at which the phase transition
occurs $T_*$, which is parameterized by the Hubble scale $\mathcal{H}_*$).
$\Om_{S*}$ and $v$ are related, in a way which depends on
the characteristics of the phase transition: for example its
strength, the properties of the bubble expansion, the interactions of
the fluid particles with the field which is undergoing the transition,
and so on.  In early works on GWs from bubble
collisions, it has been assumed that the bubbles expand as a Jouguet
detonation, because in this case the above parameters can be
calculated quite straightforwardly \cite{KKT}. However, the recent analysis of \cite{geraldine} 
demonstrates that there is no particularly justified reason for this assumption, and 
other kinds of solutions for the bubble expansion are possible, such as deflagrations, runway solutions and hybrids. Ref.~\cite{geraldine} presents a model-independent description of the different regimes characterizing the bubble expansion, including the effect of friction due to the interaction of the bubble wall with the fluid particles. From a given particle physics model, one can in principle evaluate the friction parameter $\eta$ and the strength of the phase transition $\al = \rho_{\rm vac}/\rho_{\rm rad}$. Once these two quantities are known, Ref.~\cite{geraldine} provides a way to determine the bubble wall velocity $v$ and the fraction of vacuum energy density which goes into kinetic energy of the bubble walls, $\kappa=\rho_{\rm kin}/\rho_{\rm vac}$. In terms of the parameter $\kappa$, the fractional source energy density for bubbles
becomes
$$ \Om_{S*}^{({\rm B})}= \ka \frac{\al}{\al+1} \,.$$
In the case of MHD turbulence, one further has to convert a part of the
bubble wall kinetic energy to turbulence and magnetic fields. The efficiency
of this conversion is not straightforward to estimate. Ref.~\cite{geraldine} provides a relation between the bubble wall velocity and the 
fluid velocity at the bubble wall position $v_f$: in the most optimistic case, one can argue  that the overall
kinetic energy in turbulence is simply determined by this 
fluid velocity. In this case one would have $\Om_{S*}^{({\rm MHD})}\sim v_f^2/2$.

In the absence of a way to determine $\alpha$ and $\eta$ from a given particle physics model, in the present analysis we have decided
to keep the parameters completely model-independent. We make the
simple assumption of equipartition, namely we assume the same energy
density in colliding bubble walls and in MHD turbulence, $
\Om_{S*}^{({\rm B})}= \Om_{S*}^{(\rm MHD)}$.  This is more a
reflection of our ignorance of how this energy density will be
distributed than a well justified assumption; nevertheless, it seems
to be a reasonable expectation and in the absence of a model for the
phase transition it is the most straightforward assumption.  We also
assume a strongly first order phase transition, which induces
super-sonic bubble velocities, $v>c_s$. We set the temperature of the
QCD phase transition at $T_* =100$ MeV. The other parameter relevant for the GW spectra is the duration of the phase transition, parametrised by 
$\beta$. In the electroweak case, this is usually taken to be (1 -- 10)\% of
a Hubble time: $\beta = (10 -100)\HH_*$. This value is based on the estimate given in Ref.~\cite{hogan83}, which shows that $\beta$ is related to the temperature of the phase transition through $\beta/\mathcal{H}_*\sim 4\ln(m_{\rm Pl}/T_*)$, for a phase transition nucleated via thermal fluctuations. In the absence of a precise
model for the QCD phase transition, in this section we have decided to take $\beta = 10\,\HH_*$, which is more
favorable for observations with pulsar timing experiments. The analysis of \cite{hogan83} demonstrates that models of phase transitions with small values of $\beta/\mathcal{H}_*$ may be rather exceptional, but cannot be ruled out by general arguments. 
The important point is that  $\beta/\mathcal{H}_*$ must be 
larger than unity, otherwise the phase transition is not fast with respect
to the universe expansion and our assumptions no longer hold.

\begin{figure}[ht]
\includegraphics[width=7cm]{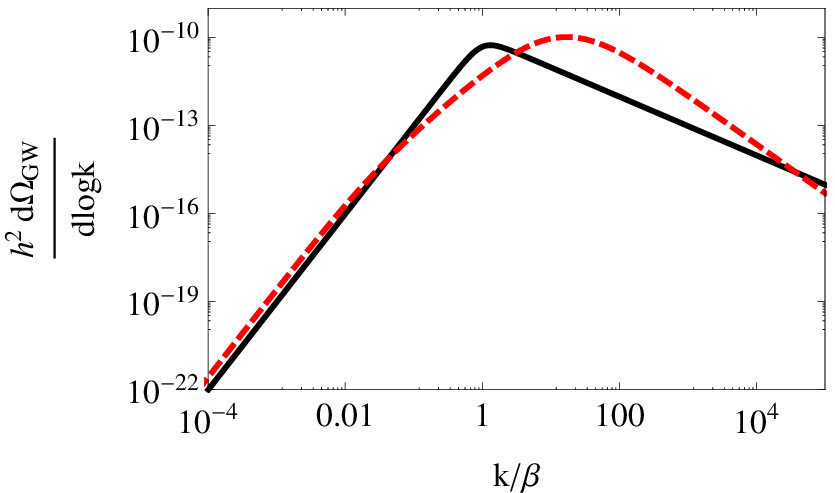}\\
\includegraphics[width=7cm]{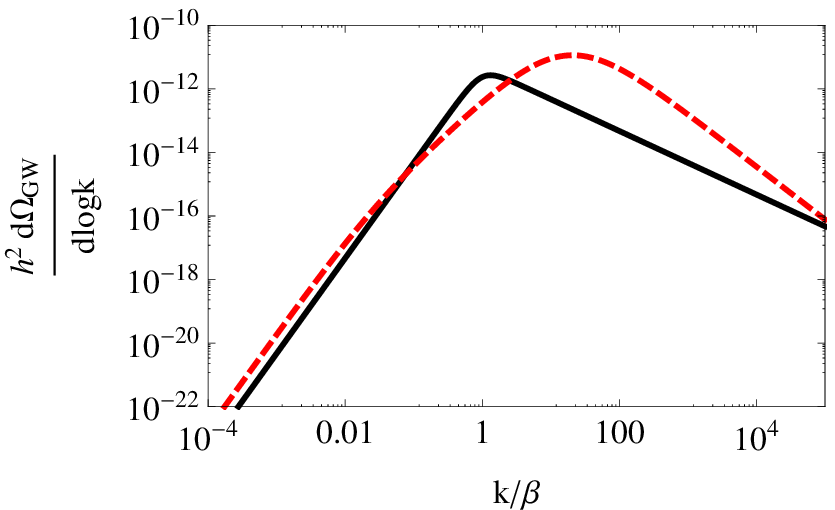}\\
\caption{\label{f:parm}The GW spectra from bubble collisions
(black, solid) and from MHD turbulence (red, dashed) are shown for different 
values of $\Om_{S*}=0.1$ and $v=0.7$ (top panel) and $\Om_{S*}=0.03$ and 
$v=0.57\simeq c_s$ (bottom panel). We set $\beta=10\,\HH_*$ and $T_*=100$ MeV 
throughout.}
\end{figure}

In Fig.~\ref{f:parm} we show the GW spectrum for both,
bubbles and MHD turbulence for two different choices of the parameters
$\Om_{S*}$ and $v$. The MHD turbulence signal dominates almost in the entire frequency
range. At large scales, it is slightly higher due to the long duration
of the turbulent source with respect to bubble collisions \cite{CRG}.
As already mentioned, the long duration of the source also causes the
slope of the MHD signal to change at sub-horizon scales from $k^3$ to
$k^2$: consequently, for $\beta>k>\mathcal{H}_*$, i.e. $0.1<k/\beta<1$, the bubble collision signal
prevails. This is valid up to the peak of the bubble collision signal,
which arises before the turbulent one: at $k/\beta\simeq 1$,
corresponding to the inverse characteristic time of the source, while
the turbulent spectrum peaks at $k/\beta\simeq \pi^2/v$, corresponding
to the inverse characteristic scale of the source. This causes the
turbulent signal to dominate at interesting frequencies, since the
total spectrum continues to raise after $k/\beta\simeq 1$ (only if the
energy in turbulence is about one order of magnitude smaller than the
one in bubble collisions, the collision signal will dominate: however,
this seems somewhat unnatural given the extremely high Reynolds number
of the primordial fluid, and we discard this possibility in this
work) \footnote{Contrary to Ref.~\cite{nicolis}, we find that the expected peak frequency of the GW spectrum from bubble collisions is always smaller than the one from MHD turbulence, the former being related to the duration of the phase transition while the latter to the size of the bubbles. This discrepancy arises because Ref.~\cite{nicolis} assumed that the peak frequency for the GW spectrum from MHD turbulence is related to the turbulent eddy turnover time, while in our case it is determined by the time correlation properties of the GW source, as explained in details in Ref.~\cite{CRG}.}.

\begin{figure}[ht]
\includegraphics[width=8.5cm]{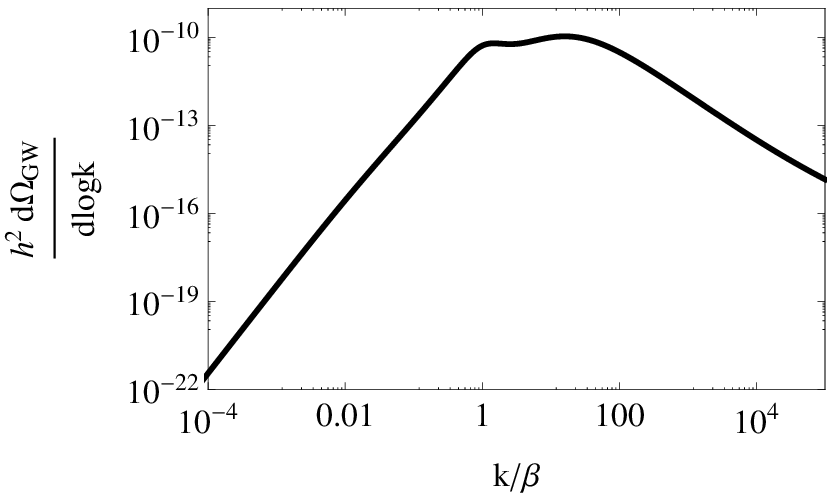}
\caption{\label{f:signal} The GW signal from bubble collisions
and MHD turbulence for $\Om_{S*}=0.1$ and $v=0.7$.
We choose $\beta=10\, \HH_*$. The signal is dominated by the 
contribution from MHD turbulence. The bubble collision peak causes the hump on the left of the true peak of the spectrum.}
\end{figure}

In Fig.~\ref{f:signal} we show the total signal for the more optimistic case,
 $\Om_{S*}=0.1$ and $v=0.7$. The peak frequency of the total GW spectrum corresponds to the MHD
turbulence peak: $k/\beta\simeq \pi^2/v$, and depends on the choice
$\beta=10\, \HH_*$. From $f=k/(2\pi)$ one obtains \cite{mybook,R}
\be
f_p\simeq 1.7 \cdot 10^{-9} \, \frac{\pi^2}{v}\,
\frac{\beta}{\HH_*}\left(\frac{g_*}{10} \right)^{\frac{1}{6}} 
\frac{T_*}{100\,{\rm MeV}}\, {\rm Hz}
\label{fp}
\ee
where $g_*$ is the number of effective relativistic degrees of freedom
at the temperature $T_*$. With $v=0.7$, $\beta=10\,\HH_*$, $g_*= 10$
and $T_*=100\,{\rm MeV}$ the peak frequency becomes $f_p\simeq
2.5\cdot 10^{-7}$ Hz.

\section{The pulsar timing array}\label{s:pul}

Neutron stars can emit powerful beams of electromagnetic waves from
their magnetic poles. As the stars rotate the beams sweep through
space like the beacon of a lighthouse. If the Earth lies within the
sweep of a neutron star's beams, the star is observed as a point
source in space emitting short, rapid pulses of 
electromagnetic waves, and is referred to as a pulsar.

\begin{figure}[ht]
\includegraphics[width=9.5cm]{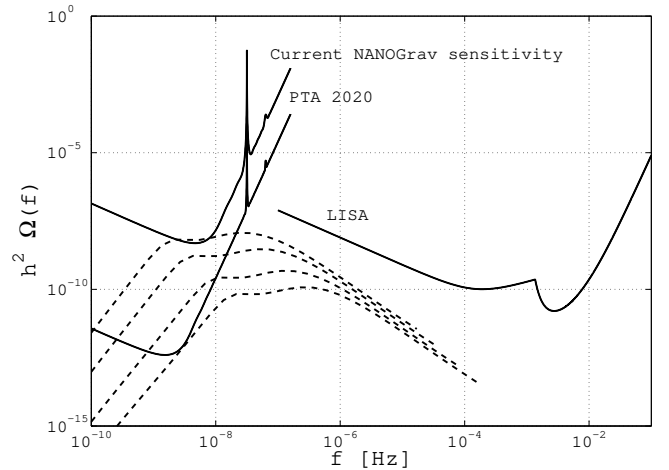}\\
\caption{
Comparison of the GW spectrum $h^2\Omega(f)$ with current NANOGrav 
pulsar timing array sensitivity and expected sensitivity of pulsar timing experiments in 
2020~\cite{Demorest:2009ex}. We have used $h=0.73$, $\Om_{r0}=8.5\times 10^{-5}$, 
$\Om_{S*}=0.1$ and $v=0.7$. We plot the GW spectra for the  values  
${\HH_*}/{\beta}=1,~0.5,~0.2,~0.1$  (dashed lines from top to bottom). 
For ${\HH_*}/{\beta} \sim 1$, the background of GWs can just be detected in 
present pulsar timing experiments, while for $0.1\lesssim {\HH_*}/{\beta}$ it can be detected by the planned array IPTA2020 (very high values of ${\HH_*}/{\beta}\sim 1$ are difficult to accommodate in the case of a thermally nucleated phase transition, c.f. discussion in the text). We also show the LISA sensitivity~\cite{lisa1,lisa2}. Unfortunately, 
LISA will not be able to detect a signal from a first order QCD phase transition (the EW phase transition is more promising in this respect \cite{Kos92,Kos93,KKT,apreda,apreda2,HuKo2,HuKo,bubble,CRTG,hogan832,Kos02,dolgov,%
CR,gogob,THelical,THelical2,CRG,R,nicolis}).}
\label{f:omega}
\end{figure}

The electromagnetic pulses we observe arrive at a very steady rate due
to the enormous moment of inertia of neutron stars.  
The idea to use these stable clocks to detect GWs
was first put forward in the late 1970s~\cite{Sazhin78,Detweiler,LK}.  
Fluctuations in the time of arrival of pulses, after all known effects are subtracted,
could be due to the presence of GWs.  Recently pulsar
timing precision has improved dramatically. Jenet and
collaborators~\cite{Jenet:2005pv} have shown that the presence of
nano-Hertz GWs could be detected using a pulsar timing
array (PTA) consisting of 20 pulsars with timing precisions of 100
nanoseconds over a period of 5 to 10 years (see
also~\cite{Hobbs:2009yy,Demorest:2009ex} for more recent PTA
sensitivity estimates). Pulsar timing arrays are most sensitive
in the band $10^{-9}$~Hz~$<f < 10^{-7}$~Hz.  The lower limit in
frequency is given by the duration of the experiment ($\sim 10$~yr.) and the upper
limit by the sampling theorem, i.e. the time between observations ($\sim 1$ month).
The spike in the sensitivity at $f=0.3\times 10^{-7}$Hz seen in Fig.~\ref{f:omega} 
is the frequency of the earth's rotation around the sun which cannot be disentangled 
from a GW with the same frequency.

\begin{figure}[ht]
\includegraphics[width=9.5cm]{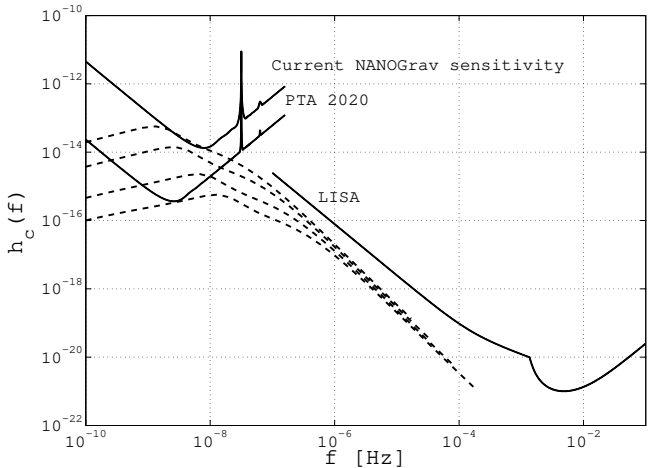}\\
\caption{Same as Fig.~\ref{f:omega} but comparing 
the characteristic strain $h_c$ as given by Eq.~(\ref{eq:hc}).}
\label{f:hc}
\end{figure}

The North American Nanohertz Observatory for Gravitational Waves (NANOGrav) \cite{nanograv}, a 
collaboration of astronomers, has created a pulsar timing array--a galactic scale 
GW observatory using about 20 pulsars. It is a section of the IPTA, an 
international collaboration involving similar organizations of European and Australian 
astronomers. The current  NANOGrav pulsar timing array sensitivity is shown in 
Fig.~\ref{f:omega}, together with the GW spectra we expect from the QCD phase transition as a function of frequency
\be
h^2\Omega_{\rm GW}(f)=h^2\frac{d\Omega_{\rm GW}}{d \log k}\,,
\ee
for ${\HH_*}/{\beta}=1$ 
(top dashed line), ${\HH_*}/{\beta}=0.5$ (upper-middle dashed line), ${\HH_*}/{\beta}=0.2$ 
(lower-middle dashed line), ${\HH_*}/{\beta}=0.1$ (bottom dashed line).
We have used
$h=0.73$, $\Om_{r0}=8.5\times 10^{-5}$ (which includes photons and neutrinos),
$\Om_{S*}=0.1$ and $v=0.7$.  
We have taken the frequency to be (c.f. Eq.~(\ref{fp}))
$$f = 1.7 \times 10^{-9} \left(\frac{\beta}{\HH_*}\right)  \left(\frac{k}{\beta}\right){\rm Hz}$$
i.e. chosen $T_*=100\,{\rm MeV}$, $g_*=10$, and $k/\beta$ varying
between $10^{-4}$ and $10^4$ (like  in Figs~\ref{f:parm} and \ref{f:signal}). The signal is compared with 
the current sensitivity of the NANOGrav pulsar timing array, and the expected
sensitivity of the IPTA pulsar timing array in 2020~\cite{Demorest:2009ex}. For values
of $0.1 \lesssim {\HH_*}/{\beta} \lesssim 1$, the background of GWs
would be detected with future pulsar timing array sensitivities. The value of ${\HH_*}/{\beta}$ must certainly be
smaller than unity for the phase transition to be fast with respect to the Hubble time and for our 
approximations to apply. In most cases, if the phase transition happens at a temperature much smaller than the Planck temperature, ${\HH_*}/{\beta}$ is of the order of 0.01; however, higher values of this parameter cannot be excluded, and we adopt them here since they are more promising for detection \cite{hogan83}. 

Fig.~\ref{f:omega} also shows the sensitivity of the planned Laser Interferometer Space Antenna 
(LISA)~\cite{lisa1} assuming that some of the confusion noise from white dwarf binaries can 
be subtracted out~\cite{lisa2}. LISA will not be able to detect 
the GW signature of a first order QCD phase transition: in order to be detectable by LISA, the GW spectrum must peak at higher frequency and consequently the phase transition must occur at higher temperature. LISA can in principle detect GWs from a strongly first order EW phase transition at $T_*\simeq 100$ GeV \cite{Kos92,Kos93,KKT,apreda,apreda2,HuKo2,HuKo,bubble,CRTG,hogan832,Kos02,dolgov,%
CR,gogob,THelical,THelical2,CRG,R,nicolis,ashorioon}.

A related quantity often used in the pulsar timing community is the (dimensionless) 
characteristic strain, defined by~\cite{MM}
\be
h_c^2 (f) =\frac{3H_0^2}{2\pi^2}f^{-2}\Omega_{\rm GW}(f) \; .
\label{eq:hc}
\ee
In Fig.~\ref{f:hc} we show the same data as in 
Fig.~\ref{f:omega} but in terms of the characteristic strain $h_c$.

\section{Conclusion and outlook}\label{s:con} 

A stochastic background of GWs from the QCD phase transition could be 
detected by pulsar timing experiments if the transition is strongly first order and lasts sufficiently long with respect to the Hubble time. In standard 
cosmology the QCD phase transition is not even second order, but simply a cross-over and in this 
case we do not expect it to generate GWs. However, if the neutrino chemical 
potential is sufficiently large~\cite{schwarz}, the QCD phase transition does become
first order. The required chemical potential does not violate nucleosynthesis constraints, and if a 
sterile neutrino is the dark matter, we do actually expect a large neutrino chemical 
potential~\cite{misha}. 

Pulsar timing experiments will reach unprecedented sensitivities in the next few years, and 
may open a new window on cosmology. The detection of a stochastic background with pulsar 
timing experiments could help to study the nature of the QCD phase transition, its duration,
its strength an so on; by comparison with lattice calculations, this would allow us to determine the 
neutrino chemical potential and other properties of the so elusive cosmological neutrino sector. 
Furthermore, the amplitude and peak frequency of the spectrum are sensitive to the 
expansion rate of the Universe at this temperature~\cite{chuang}, which remains unconstrained to 
date. 

A first order QCD phase transition generating a GW background would also
induce a stochastic background of magnetic fields, as studied 
in the past~\cite{magQCD}. Furthermore, the GW background might have non-vanishing 
helicity, which would be an interesting phenomenon to investigate by itself~\cite{elisa}.

\begin{acknowledgments} 
  We would like to thank Paul Demorest, Andrea Lommen, Larry Price and Geraldine 
  Servant for useful conversations.  RD acknowledges support from the Swiss National 
  Science Foundation. XS is supported in part by NSF Grant No.
  PHY-0758155 and the Research Growth Initiative at the University of
  Wisconsin-Milwaukee.
\end{acknowledgments}

\end{document}